\documentstyle[12pt]{article}
\newcommand{\bq}{\begin{equation}}
\newcommand{\ee}{\end{equation}}
\newcommand{\bi}[1]{\bibitem{#1}}
\newcommand{\fr}[2]{\frac{#1}{#2}}
\newcommand{\etab}{\overline\eta}

\begin{document}
\pagestyle{plain}
\pagenumbering{arabic}

\begin{flushright}
{\large Preprint BUDKERINP 94-70}
\end{flushright}

\begin{center}{\Large \bf Instanton--Antiinstanton pair
induced Asymptotics of Perturbation Theory in QCD }\\

\vspace{0.5cm}

{\bf  P.G.Silvestrov}\\
Budker Institute of Nuclear Physics, 630090 Novosibirsk,
Russia

\vspace{0.5cm}

\end{center}
\begin{abstract}

The instanton--antiinstanton pair induced asymptotics of
perturbation theory expansion for QCD correlators is
considered. It is argued that though the true asymptotics
is dominated by renormalon, the instanton-induced
contribution may dominate in the intermediate asymptotics
$n= 5\div 15$.

Obtained asymptotic formulae are valid for $N_f \le N_c$.
For $N_f =N_c$ the finite nonperturbative expression for
instanton--antiinstanton contribution was also found.

At $q^2<0$ the imaginary part of correlators in the case
$N_f =N_c$ is suppressed like $1/\log (q/\Lambda)$, but the
present accuracy of instanton calculations allows to fix it
unambiguously.

The series of corrections to the instanton induced
asymptotics of the order of $\sim (\log(n)/n)^k$ is found.

\end{abstract}

\section{Introduction. Renormalon--Instanton}\label{sec:1}

The aim of this note is to try to show, what the important
role may play the instanton for large order terms of
perturbation theory in QCD. During last few years a lot of
papers appears (see e.g.
\cite{Muel,West,Brow,VZre,Bene,Grun,Broa})
considering the asymptotic behaviour of
perturbation theory series in QCD and QED.  However the
main attention was paid to the so called renormalon
asymptotics.  Two renormalons were considered, the
ultraviolet and infrared.  The usual form of the
ultraviolet renormalon contribution to, for example,
$R_{e^+e^-
\rightarrow hadrons}$ perturbative expansion is
\bq\label{eq:uren}
a_n \sim \left( -b \fr{\alpha_s}{4\pi}\right)^n n! \,\,\, ,
\ee
where $b =11/3 N_c -2/3 N_f\approx 10$. In QCD the series
(\ref{eq:uren}) is sign alternating and at least the Borel
sum of the series is well defined. The problems with
summation of ultraviolet renormalon appears in QED, where
all terms of series have the same sign.

Another kind of asymptotics is the infrared renormalon:
\bq\label{eq:iren}
a_n \sim \left( \fr{b}{2} \, \fr{\alpha_s}{4\pi}\right)^n
n! \,\,\, .
\ee
At large $n$ this series turns to be much smaller than
(\ref{eq:uren}). The keen interest in infrared renormalon
was caused by the fact that nobody knows how to sum the
series (\ref{eq:iren}). The Borel sum of this series is ill
defined (up to arbitrary $\sim 1/q^4$ correction).

Both ultraviolet and infrared renormalons are associated
with a certain chains of Feynman graphs. However since the
works of L.N.Lipatov
\cite{Lipatov} another approach is developed for large
order perturbation theory estimates. In this approach one
has nothing to do with Feynman graphs, but tries to find
the specific ("classical") large fluctuations in the
functional space making the main contribution to the high
order terms of perturbative expansion. The natural example
of such important fluctuations in QCD is the
instanton--antiinstanton pair,but up to now only in the
paper of I.I.Balitsky \cite{Balitsky} the instanton
asymptotics for $R_{e^+e^-\rightarrow hadrons}$ was
considered. In present work we try to clarify further the
role of instanton--antiinstanton pair effects for QCD
correlators, especially in the interesting case $N_f=N_c$
(the author of
\cite{Balitsky} at $N_f=N_c$ have not found the
perturbative asymptotics and use the ambiguous
regularization prescription to find nonperturbative
corrections).

The generic form of instanton--induced asymptotics appears
to be:
\bq\label{eq:insas}
a_{nIA}\sim \left(\fr{\alpha_s}{4\pi}\right)^n (n+4N_c)!
\,\,\, .
\ee
The overall numerical factor in $a_{nIA}$ may also be
sufficiently large. It is seen that though the renormalons
(\ref{eq:uren},\ref{eq:iren}) do dominate at very large $n$
($n>15$), the instanton-induced contribution may dominate
in the intermediate asymptotics $n= 5\div 15$. If so, the
pure renormalon behaviour (\ref{eq:uren}) will hardly be
observed in directly calculated terms of perturbation
theory due to a strong competition with the instanton
contribution.

Of course the exactly known $3\div4$ terms of perturbation
theory series (for $\beta$--function of QCD or
$R_{e^+e^-\rightarrow hadrons}$) are much smaller than the
estimate (\ref{eq:insas}) and the question at what number
$n$ the perturbative series could reach the full strength
(\ref{eq:insas}) is open now. Moreover the corrections
which are formally $\sim 1/n$ at large $n$ may also change
the value of $a_n$ in many orders at small $n$. For example
\bq\label{eq:Lip}
(n+4N_C)! =  n^{4N_C} n! (1+O(1/n)) \,\,\, .
\ee
In this paper we have also found (and summed up) the
subseries of corrections to $a_{nIA}$ (\ref{eq:insas}),
which behave like $(\log(n)/n)^k$ (see
(\ref{eq:asympcor})).

Like for the infrared renormalon (\ref{eq:iren}) all terms
of the series (\ref{eq:insas}) have the same sign.
Nevertheless the problem of summation the series
(\ref{eq:insas}) seems not so hopeless as the summation of
renormalon. Following G.'t Hooft \cite{t'Hooft} the author
of
\cite{Balitsky} proposed to rewrite the integral over the
instanton--antiinstanton pair in the Borel form by
considering the action as a collective variable. The
well--separated instanton--antiinstanton pair is
responsible for the singular part of Borel function, while
the ambiguous strongly interacting instanton and
antiinstanton contribute to its smooth part. On the other
hand the best way to describe the smooth part of the Borel
transform is to calculate exactly the few first terms of
perturbative expansion. The divergent singular part of the
Borel integral corresponds to almost non-interacting
pseudoparticles. The accurate subtraction from the
singularity of dilute gas contribution in the toy model
(double well oscillator) allowed to find the finite
nonperturbative instanton--antiinstanton contribution
\cite{Faleev}. In QCD at $N_f=N_c$ the Borel integral
diverges only logarithmically and the total nonperturbative
contribution from instanton--antiinstanton pair may be
found by cutting the instanton size at $\rho \ll
1/\Lambda$.

At $N_f=N_c$ the imaginary part of correlators at negative
$q^2$ cancels in the one loop approximation. Nevertheless
the invariance of the instanton contribution under the
renormalization group transformations allows to find the
imaginary parts.

\section{The ansatz}\label{sec:2}

The most popular puzzle for applying the high order
estimates is the calculation of $R_{e^+e^-\rightarrow
hadrons}$, which is connected with the Euclidean correlator
of two electromagnetic currents
\bq\label{eq:emcorr}
\Pi_{\mu\nu} = \int e^{iqx} d^4x <j_{\mu}(x)j_{\nu}(0)>
\,\,\, ,
\ee
where $j_{\mu}=\sum_{flavours} e_f \Psi_f^+ \gamma_{\mu}
\Psi_f$.  Calculation of instanton--induced contribution to
(\ref{eq:emcorr}) requires considerable algebraic efforts
(e.g. the fermionic Green function in the pseudoparticle
background must be used). Therefore for the sake of
simplicity we will examine the correlator of two scalar
currents
\bq\label{eq:current}
j(x)= \fr{3\alpha_s}{4\pi}\left[ G^a_{\mu\nu}(x)\right]^2
\ee
(the notations of \cite{NSVZ} are used). This correlator,
which may be useful for the gluebal physics, is simple
enough so the reader can check almost all steps of the
calculation. Moreover the correlator of two currents
(\ref{eq:current}) reproduces all the interesting features
of the correlator of electromagnetic currents.

As we have said above, the strongly interacting instanton
and antiinstanton correspond to a smooth part of the Borel
function. It was shown by Balitsky
\cite{Balitsky} that the instanton--antiinstanton
configuration relevant for the large orders of perturbation
theory is a small instanton inside the very large
antiinstanton (or vice versa). The size of small instanton
is regulated by the internal momentum of correlator
(\ref{eq:emcorr}) $q
\rho_I \sim 1$. The size of antiinstanton (as well as the
distance between the pseudoparticles $\rho_A \sim R$)
becomes parametrically large as we consider the higher
terms of perturbation theory.

Now let us specify the ansatz for the gauge fields. We are
interesting in the instanton--antiinstanton interaction in
the leading nontrivial approximation. Therefore the simple
sum of instanton and antiinstanton may be used:
\bq\label{eq:ansatz}
A_{\mu} = U_A A_{\mu}^A U_A^+ + U_I A_{\mu}^I U_I^+ \,\,\,
,
\ee
where $U_A, U_I$ are the constant $SU(N)$ matrices. For
small instanton the singular gauge seems to be preferable
\bq\label{eq:Ains}
A_{\mu}^I = \fr{\etab_{\mu\nu}(x-x_I)_{\nu} \rho_I^2}
{(x-x_I)^2((x-x_I)^2+\rho_I^2)} \,\,\, , \,\,\,
\etab_{\mu\nu}\equiv \tau^a \etab^a_{\mu\nu} \,\,\, .
\ee
Before we add the antiinstanton to (\ref{eq:Ains}) the
singularity at $x=x_I$ is pure gauge singularity. Therefore
in order to remove the singularities from all the physical
quantities one may choose $A_{\mu}^A$ in any regular gauge
which satisfy the equality $A_{\mu}^A(x=x_I)=0$. For
example one can rotate the BPST antiinstanton
\begin{eqnarray}\label{eq:Aans}
A_{\mu}^A = S\left[ \fr{\etab_{\mu\nu}(x-x_A)_{\nu}}
{(x-x_A)^2+\rho_A^2}
\right] S^+ + i S\partial_{\mu} S^+\,\,\, , \\
R_{\mu}= (x_A-x_I)_{\mu} \,\,\, , \,\,\,
S= \exp\left\{ i\fr{\etab_{\mu\nu} R_{\nu}}{R^2+\rho_A^2}
(x-x_I)_{\mu}\right\} \,\,\, . \nonumber
\end{eqnarray}
It is easy to show that any other smooth matrix function
$S(x)$, which allows to cancel the antiinstanton field at
$x=x_I$ will lead to the same correlator.

After direct calculation the classical action of the
instanton--antiinstanton configuration may be found with
the usual dipole--dipole interaction of pseudoparticles
\bq\label{eq:action}
S_{IA} = \fr{4\pi}{\alpha_s}\left\{
1 - \xi h\right\} \,\,\,\, , \,\,\,
\xi = \fr{\rho_I^2 \rho_A^2}{(R^2 + \rho_A^2)^2} \,\,\, ,
\,\,\,
h=2|TrO|^2 -TrOO^+ \,\,\, ,
\ee
and $O$ is the upper left $2 \times 2$ corner of the matrix
$U=U_A^+ U_I$ (\ref{eq:ansatz}).

Another part of the problem, extremely sensitive to the
instanton--antiinstanton interaction, is the fermionic
determinant. It may be shown, that for each flavor of
massless fermions the two anomalously small eigenvalues of
Dirac operator $\hat{D}$ appears
\bq\label{eq:lambda}
\lambda_{1,2}= \pm \fr{2\rho_I \rho_A}{(\rho_A^2 +
R^2)^{3/2}}|TrO| \,\,\, .
\ee

\section{Calculation of correlator}\label{sec:3}

After we have defined the gauge field configuration it is
easy to write down the instanton--antiinstanton
contribution to the correlator of two currents
(\ref{eq:current}). Everywhere it is possible the notations
of \cite{Balitsky} are used.
\begin{eqnarray}\label{eq:corr}
&\,&\Pi (q)=\int e^{iqx} d^4x <j(x)j(0)> = \\
&\,&=2 \int j_I(x) j_I(0) e^{iqx} [4 \xi^{3/2}
|TrO|^2]^{N_f}
\exp \left\{ \fr{4\pi}{\alpha_s}\xi h\right\}
\fr{d(\rho_I)}{\rho_I^5}
\fr{d(\rho_A)}{\rho_A^5} dx dx_I dx_A d\rho_I d\rho_A dU
\,\, ,
\nonumber
\end{eqnarray}
where
\bq\label{eq:xi}
j_I(x)=\fr{36}{\pi^2} \fr{\rho^4}{((x-x_I)^2+\rho_I^2)^4}
\,\,\, ,
\ee
and the instanton density \cite{Bernard}
\begin{eqnarray}\label{eq:dins}
d(\rho)= A \left( \fr{2\pi}{\alpha_s(\rho)}\right)^{2N_c}
\exp\left(-\fr{2\pi}{\alpha_s(\rho)} \right)  \,\,\, .
\end{eqnarray}
The factor $2$ in front of the integral in (\ref{eq:corr})
accounts for the equal contribution from small
antiinstanton and large instanton.

We will also use the well known two--loop formula
\bq\label{eq:alpha}
\fr{4\pi}{\alpha_s(q)} = b\log
\left(\fr{q^2}{\Lambda^2}\right)+
\fr{b'}{b}\log\left(
\log\left(\fr{q^2}{\Lambda^2}\right)\right) + ...
\,\,\, .
\ee
In the most interesting case $N_f=N_c=3 \,\,\,\,\, b=9$ and
$b'=64$.

Before passing to the formal computations let us say a few
words about the existence of integral (\ref{eq:corr}) as a
whole. The most ambiguous part of the problem is the
integration over large antiinstanton coordinates $\rho_A$
and $R=x_A-x_I$. There are two competing effects. The
factor $d(\rho_A) \sim \rho_A^{b}$ tends to make the
integral over $\rho_A$ divergent. On the other hand, the
almost zero fermionic modes (\ref{eq:lambda}) tends to
suppress the large $\rho_A$ and $R$ contribution.  If
$N_f\le N_c$ the first effect dominates and the integral
(\ref{eq:corr}) diverges at large $\rho_A$. Nevertheless
just in this case the well defined instanton induced
asymptotics of perturbation theory may be extracted from
(\ref{eq:corr}). For nonperturbative calculation of the
whole integral (\ref{eq:corr}) at $N_f\le N_c$ the new
physical income is necessary. Below we will show how to
perform this integration for $N_f = N_c =3$.

If the number of massless flavours is sufficiently large
($N_f > N_c$) the attraction due to fermionic zero modes
prevails. As a result the approximation of almost
noninteracting pseudoparticles breaks down ($\rho_A
\sim R \sim \rho_I$) and the instantonic approach itself
became ambiguous.

Formulae (\ref{eq:dins}), (\ref{eq:alpha}) allow to extract
the $\rho_A$ dependent part from (\ref{eq:corr})
\bq\label{eq:phi}
d(\rho_A) = \phi (\rho_I^2 / \rho_A^2 )
 \left( \fr{\rho_I}{\rho_A}
\right)^b d(\rho_I) \,\,\,\, , \,\,\,\,
\phi(x) = \left[ 1 +
b\fr{\alpha_s}{4\pi}\log (x)\right]^{2N_c-\fr{b'}{2b}}
\ee
Everywhere below we suppose $\alpha_s = \alpha_s(q) \simeq
\alpha_s (\rho_I)$. For calculation of leading perturbative
asymptotics one may assume $\phi(x)\equiv 1$ (as it was
done in \cite{Balitsky}), but in order to calculate the
nonperturbative value of correlator (\ref{eq:corr}) the
function $\phi(x)$ of the form (\ref{eq:phi}) should be
used. Moreover the corrections $\sim (\alpha_s \log(x))^k$
contained in the $\phi (x)$ leads to an important subseries
of preasymptotic corrections $\sim (\log(n)/n)^k$ to the
leading asymptotics of the perturbative expansion.
Therefore below we shell use the function $\phi(x)$, though
suppose that its argument is small enough $|\log(x)|\gg 1$.

Now we would like to integrate over $\rho_A$ and
$R=x_A-x_I$ for a fixed value of $\xi$
\bq\label{eq:rhoA}
\int \phi (\rho_I^2 / \rho_A^2 ) \rho_A^{b-5} \delta \left(
\fr{\rho_I^2\rho_A^2}{(R^2+\rho_A^2)^2}-\xi\right) d\rho_A
d^4R =
\fr{\pi^2}{2(b-2)(b-1)} \fr{\rho_I^b}{\xi^{b/2+1}}
\phi(\xi) \,\,\, .
\ee
The part of (\ref{eq:corr}) depending on $\rho_I,x$ and
$x_I$ in the leading approximation over $\alpha_s$ gives:
\begin{eqnarray}\label{eq:int}
\int j_I(x) j_I(0) e^{iqx} \rho^{2b-5} d^4x d^4x_I d\rho =
9
\fr{ 2^{2b-3}}{q^{2b-4}} \fr{\Gamma(b+2) \Gamma^2(b)
\Gamma(b-2)}
{\Gamma(2b)} \,\,\, .
\end{eqnarray}

After all the correlator (\ref{eq:corr}) reads
\begin{eqnarray}\label{eq:last}
\Pi(q) = Const \,\, q^4
d^2(1/q) \int |TrO|^{2N_f}
\exp \left\{ \fr{4\pi}{\alpha_s}\xi h\right\}
\fr{\phi(\xi)d\xi}{\xi^{\fr{b-3N_f}{2}+1}} dU \,\,\, .
\end{eqnarray}

The last step, which allows to rewrite the correlator in
the form of Borel integral is to introduce the variable
$t=1- \xi h$:
\begin{eqnarray}\label{eq:corrBor}
\Pi(q)&=& 9\pi^2 2^{2(b+N_f-2N_c)-3}
\fr{(b+1)!(b-1)!(b-3)!^2}{(2b-1)!}
A^2 q^4 \left(\fr{4\pi}{\alpha_s(q)}\right)^{4N_c} \\
&\,&\left\{ \int_0^1 dt\phi(1-t) (1-t)^{\fr{3N_f-b}{2}-1}
e^{-\fr{4\pi}{\alpha_s}t}
<|TrO|^{2N_f} \Theta(h) h^{\fr{b-3N_f}{2}}> \nonumber +
\right. \\
&+& \left.
\int_1^{\infty} dt \phi(t-1) (t-1)^{\fr{3N_f-b}{2}-1}
e^{-\fr{4\pi}{\alpha_s}t}
<|TrO|^{2N_f} \Theta(-h) (-h)^{\fr{b-3N_f}{2}}> \right\}
\,\, . \nonumber
\end{eqnarray}
Here $\Theta$--function equals $0$ or $1$ in accordance
with sign of its argument and $<\,\, ...\,\, >$ means
averaging over the orientation of the matrix $U$.

Essentially the same expression as (\ref{eq:corrBor})
(except for the overall power of $q$ and numerical factors
and with $\phi$ replaced by $1$) was found in
\cite{Balitsky} for the correlator of electromagnetic
currents (\ref{eq:emcorr}).

\section{Analyzing the result}\label{sec:4}

The first conclusion which is to be done is that the result
(\ref{eq:corrBor}) may be used only for $N_f \le N_c$
because the integral over orientations of the matrix $U$
diverges at $h=0$ if $N_f>N_c$.  This means that our method
can not be applied without strong modification , for
example, to calculation of $\Gamma_{Z_0 \rightarrow
hadrons}$ ($N_c=3\,\, ,
\,\, N_f=5$).

Another interesting application for asymptotic formulae
will be the case $N_f=N_c=3$. The averaging over $U$ for
$SU(3)$ group gives
\begin{eqnarray}\label{eq:group}
<|TrO|^6> &=&\fr{7}{5} \,\,\, , \nonumber \\
<|TrO|^6 \Theta(h) > =1.37 \,\,\, &;& \,\,\,
<|TrO|^6 \Theta(-h) > =0.03 \,\,\, .
\end{eqnarray}
Here the averages with $\Theta$--function are the numerical
estimates. The accuracy of the last (small) value is
expected to be not worse than $20\%$.

Thus the final expression for instanton--antiinstanton pair
contribution at $N_f=N_c=3$ reads
\begin{eqnarray}\label{eq:corrSU3}
\Pi (q)&=&\fr{9}{\pi^2}e^{5/3}\fr{10!7!6!^2}{17!}
 q^4 \left(\fr{4\pi}{\alpha_s}\right)^{12} \\
 & &
\left\{ 1.37\int_0^1 \phi(1-t)
\fr{\exp(-\fr{4\pi}{\alpha_s}t)}{1-t} dt
+ 0.03 \int_1^{\infty} \phi(t-1)
\fr{\exp(-\fr{4\pi}{\alpha_s}t)}{t-1} dt
\right\} \,\,\, . \nonumber
\end{eqnarray}
This expression is enough to find the leading asymptotics
of perturbative expansion for $\Pi (q)$:
\bq\label{eq:asymp}
\Pi (q)=q^4\sum \Pi_n \left(\fr{4\pi}{\alpha_s}\right)^n
\,\,\, , \,\,\,
\Pi_n = \fr{9}{\pi^2}e^{5/3}\fr{10!7!6!^2}{17!} 1.37
(n+12)! \,\,\, .
\ee
We see that instanton--antiinstanton induced contribution
to the series of perturbation theory at $n\sim 10$ do has a
huge enhancement $(n+12)!$ compared with $b^n n!$ for
renormalon. The complete calculation of the $\sim 1/n$
corrections to the leading asymptotics requires a
considerable efforts even in the simple toy model
\cite{Faleev}. Nevertheless one may try to find any
particular corrections which are enhanced in some way. The
set of such enhanced $\sim \log(n)/n$ corrections appears
from the expansion of $\phi(1-t)$ in (\ref{eq:corrSU3}) in
powers of $\alpha_s$. Let us remind, that physically with
$\phi(x)$ one takes into account the running of the
coupling $\alpha_s (\rho_A)$ which describes the large
antiinstanton. It seems very unlikely if one can find any
other effect which lead to such a large corrections $\sim
(\log(n)/n)^k$. Under this assumption we find:
\bq\label{eq:asympcor}
\Pi_n = 176 \left[ 1-9\fr{\log(n)}{n}\right]^{22/9} (n+12)!
\,\,\, .
\ee
Of course this result may be used only if $n \gg \log(n)$.
Though the expression (\ref{eq:corrSU3}) provides us with
the asymptotic of perturbative series, both integrals in
(\ref{eq:corrSU3}) diverge at $t=1$.  In the configuration
space these divergences are related to the integration over
almost noninteracting instanton and antiinstanton. Because
the divergence is only logarithmic one can try to use the
physical intuition in order to restrict the range of
integration in (\ref{eq:corrSU3}). Anyway the natural cut
for $\rho_A$ seems to be $\rho_A \ll 1/\Lambda_{QCD}$, or
in terms of $t$
\bq\label{eq:cut}
|t-1|_{min} \sim \left(\fr{\rho_I}{\rho_{Amax}}\right)^2
\ll \fr{\Lambda^2}{q^2} \,\,\, .
\ee
If so the nonperturbative part of (\ref{eq:corrSU3}) may be
found explicitly (up to corrections $\sim\alpha_s$).
\bq\label{eq:nonpert}
\Pi (q)=129 q^4 \left(\fr{4\pi}{\alpha_s}\right)^{12}
\left\{ 1.37 P \int_0^{\infty} \phi(|1-t|)
\fr{\exp(-\fr{4\pi}{\alpha_s}t)}{1-t} dt
+ \fr{7}{155} \left(\fr{4\pi}{\alpha_s}\right)
\exp(-\fr{4\pi}{\alpha_s})
\right\} \,\,\, .
\ee
Here $P$ means the principal value integral. Let us stress
that if one replaces $\phi(x)$ by $1$ in
(\ref{eq:corrSU3}), the nonperturbative part of
(\ref{eq:nonpert}) would be 31/9 times larger. Effectively
the integration over $t$ in
(\ref{eq:corrSU3}),(\ref{eq:nonpert}) may be thought as the
integration over the size of large antiinstanton. The size
of antiinstanton also may be determined through the
coupling constant $\alpha_s(\rho_A)$ (the logarithmic scale
$\alpha_s(\rho_A)^{-1} \sim -\log (\rho_A \Lambda)$). The
remarkable feature of our result  is that all the values of
$\alpha_s(\rho_A)$ contribute to the nonperturbative part
(\ref{eq:nonpert}) in the whole range $\alpha_s (\rho_I) <
\alpha_s (\rho_A) \ll 1$.

\section{Imaginary part}\label{sec:5}

In order to find the physical quantities such as the
inclusive widths and cross--sections one have to consider
the imaginary part of correlators, which appears after the
analytic continuation to Minkovsky momentum $Im(\Pi(-q^2
+i\varepsilon ))$. In the lowest order in $\alpha_s$ both
the singular part and the nonperturbative corrections in
(\ref{eq:nonpert}) behave like $(\Lambda/q)^{2b}$. Thus for
$N_f=N_c=3$ ($b=9$) the imaginary part cancels.  In this
case the second term of the expansion of $\alpha_s$ should
be considered (\ref{eq:alpha}) and the analytic
continuation of (\ref{eq:nonpert}) gives:
\begin{eqnarray}\label{eq:impart}
&\,&\fr{1}{q^4} Im\Bigl(\Pi(-q^2+i\varepsilon )\Bigr) = \\
&=& 2.4\, 10^4
\left(\fr{4\pi}{\alpha_s}\right)^{11}
P \int_0^{\infty} \psi(|1-t|)
\fr{\exp(-\fr{4\pi}{\alpha_s}t)}{1-t} dt
+ 0.9 \, 10^3 \left(\fr{4\pi}{\alpha_s}\right)^{12}
\exp(-\fr{4\pi}{\alpha_s})
 \,\,\, , \nonumber \\
 &\,& \psi(x) = \left( 1+ \fr{43}{6} \fr{4\pi}{\alpha_s}
\log(x)\right)
 \left( 1+ 9 \fr{4\pi}{\alpha_s} \log(x)\right)^{13/9}
 \nonumber
\end{eqnarray}
Very similar expression may
be found for $R_{e^+e^-\rightarrow hadrons}$. Although the
considerable
additional efforts are necessary for this calculation.

\end{document}